\documentclass[10pt, oneside,twocolumn]{revtex4}

\usepackage{amsmath,amssymb,amsfonts,verbatim,graphicx,makeidx}

\usepackage[english]{babel}

\begin{document}

\title{Procrustean entanglement concentration of continuous variable states of light}

\author{David Menzies and Natalia Korolkova}
\email{nvk@st-andrews.ac.uk}
\affiliation{School of Physics and Astronomy, University of St. Andrews, North Haugh, St. Andrews KY16 9SS, UK.}
\date{\today}

\begin{abstract}
We propose a Procrustean entanglement concentration scheme for continuous variable states inspired by the scheme proposed in [J. Fiur$\grave{a}$$\check{s}$ek {\it et. al.}, Phys. Rev. A {\bf 67}, 022304 (2003)]. We show that the eight-port homodyne measurement of [J. Fiur$\grave{a}$$\check{s}$ek {\it et. al.}, Phys. Rev. A {\bf 67}, 022304 (2003)] can be replaced by a balanced homodyne measurement with the advantage of providing a success criterion that allows Alice and Bob to determine if entanglement concentration was achieved. In addition, it facilitates a straightforward experimental implementation.
\end{abstract}

\maketitle

\section{Introduction}

High quality implementation of protocols such as quantum teleportation, super-dense coding and entanglement swapping require maximally entangled pure states \cite{nielsen&chuang:2000,plenio:1998,hirvensalo:2003,bruss:2002}. Unfortunately, processes such as decoherence and dissipation mean that maximally entangled states are potentially difficult to generate and then maintain \cite{nielsen&chuang:2000,hirvensalo:2003,plenio:1998,popescu:2000}. In a continuous variable (CV) setting, one faces the additional problem that the maximally entangled states, the so-called EPR states, are unphysical and therefore unobtainable \cite{epr:1935,leonhardt:1997,braunstein:2003}. Such pure states correspond to simultaneous eigenstates of the operators \cite{braunstein:2003,duan:2000}:
\begin{equation}
\hat{U} = \vert a \vert \hat{x}_1 + \frac{1}{a} \hat{x}_2, \; \; \; \hat{V}= \vert a \vert \hat{p}_1 - \frac{1}{a} \hat{p}_2,
\label{ops}
\end{equation}

\noindent where $\{ \hat{x}_1, \hat{x}_2, \hat{p}_1, \hat{p}_2 \}$ are the position and momentum operators of the subsystems $1$ and $2$. Consequently, the quality of any protocol using entangled CV states is always limited by the physical mechanism producing these states.

This is where entanglement concentration/distillation protocols come in, with the aim to convert a large number of copies of weakly entangled states into a small number of highly entangled states via local operations and classical communication (LOCC) \cite{bruss:2002}. There have been many different methods proposed to achieve this entanglement concentration \cite{bennett:1995}, however only the Procrustean method seems to work for CV states. In this method the {\it Schmidt} coefficients of the input state are modified whilst preserving the {\it Schmidt} basis. Symbolically, this procedure may be expressed as:
\begin{equation}
\vert \psi_{in} \rangle^{\otimes N} \stackrel{\mbox{\small LOCC}}{\longrightarrow} \vert \psi_{out} \rangle^{\otimes M},
\end{equation}

\noindent where $N \gg M$ and:
\begin{eqnarray}
\label{in}
\vert \psi_{in} \rangle = \sum_{k = 1}^{K} s_{k} \vert e_{k} \rangle \otimes \vert f_{k} \rangle, \\
\label{out}
\vert \psi_{out} \rangle = \sum_{k = 1}^{K} s'_{k} \vert e_{k} \rangle \otimes \vert f_{k} \rangle,
\end{eqnarray}

\noindent such that the entanglement content of (\ref{out}) is greater that the entanglement content of (\ref{in}).

In quantum optics, one readily obtains continuous variable entangled states from non-degenerate parametric down conversion \cite{leonhardt:1997,braunstein:2003,barnett&radmore:1997, mandel&wolf1995,loudon1973,paris:2005}, called two mode squeezed vacuum states (TMSS). Such states have the following {\it Schmidt decomposition}:
\begin{equation}
\vert \zeta \rangle = \frac{1}{\cosh r} \sum_{n=0}^{\infty} (-\tanh r)^n \vert n \rangle \otimes \vert n \rangle,
\label{squeezed}
\end{equation}

\noindent where $r $ is the squeezing parameter. Furthermore, it is useful to introduce $\lambda = \tanh r$ which allows a more convenient representation of (\ref{squeezed}):
\begin{equation}
\vert \zeta \rangle = (1- \lambda^2)^{1/2} \sum_{n=0}^{\infty} (-\lambda )^n \vert n , n \rangle = \sum_{n=0}^{\infty} c_n \vert n, n \rangle.
\end{equation}

\noindent These two mode squeezed states allow non-local correlations in the uncertainties of the quadratures of each mode but they do not correspond to the maximally entangled EPR states. However, they do asymptotically coincide with these states in the limit $r \rightarrow \infty$ \cite{alessandrini:2004,braunstein:2003}; a fact that can be easily demonstrated via the {\it Duan} separability criterion \cite{duan:1999}, where entangled states obey the condition:
\begin{equation}
\Bigl \vert a^2 - \frac{1}{a^2}  \Bigr \vert \leq \langle (\Delta \hat{U})^2 \rangle + \langle (\Delta \hat{V})^2 \rangle < a^2 + \frac{1}{a^2}.
\label{duan}
\end{equation}

\noindent Applying this to (\ref{squeezed}) (with $a=-1$) gives:
\begin{equation}
0 \leq e^{- 2 r} < 1,
\end{equation}

\noindent where the lower bound of the above is satisfied only by the maximally entangled states.

The goal here is entanglement concentration of the two mode squeezed vacuum state $\vert \zeta \rangle$ via the Procrustean method, which will result in the modification $\lambda \rightarrow \lambda'$ where $\lambda < \lambda'$. Entanglement concentration on Gaussian states can only be achieved by implementing some non-Gaussian operation in addition to local Gaussian operations and classical communication, a result that was proved in the no-go theorem provided by {\it Eisert et al} \cite{eisert:2002}. Consequently, any Procrustean CV entanglement concentration scheme must include a non-Gaussian operation to be successful. Three possible schemes have been suggested so far, the first requires quantum non-demolition measurements of photon number \cite{duan:2000} and the second utilizes single photon subtraction measurements \cite{browne:2003,eisert:2004}  while the third employs a non-linear medium \cite{fiurasek:2003}.

Here we propose a modification of a Procrustean entanglement concentration scheme suggested in \cite{fiurasek:2003}. In the setup of \cite{fiurasek:2003}, Alice and Bob initially share $\vert \zeta \rangle$ before an additional auxiliary coherent state $\vert \alpha \rangle $ is introduced. The coherent state and Bob's half of the squeezed state are fed into a nonlinear medium that exhibits the Kerr effect. This interaction results in entanglement between the coherent beam and Alice and Bob's beams. A local eight port homodyne measurement is then performed on the coherent state, which ultimately projects onto a random coherent state $\vert \beta \rangle$. However, this measurement process does not allow the construction of a success criterion, instead  all that one can say is that the entanglement content of the output state may have increased if the output $\vert \beta \rangle$ was within a certain range.  Finally, a feed forward phase shift is required in Bob's beam to remove any undesirable oscillatory terms. In our scheme, we show that by measuring one of the quadratures of $\vert \alpha \rangle$, via balanced homodyne detection, we can also have Procrustean entanglement concentration that has the added advantage of producing a success criterion. Furthermore, the experimental implementation of the homodyne detection is considerately simpler to execute that the eight port homodyne detection.

\section{The protocol}

Initially, we assume that Alice and Bob share the state $\vert \zeta \rangle$ and we introduce an additional mode $C$ prepared in a coherent state $\vert \alpha \rangle$ with $\alpha \in \mathbb{R}$. The input state to the protocol is then $\vert \psi_{in}(0) \rangle = \vert \zeta \rangle \vert \alpha \rangle$. Both Bob's beam and the additional coherent state are fed into a non-linear medium exhibiting the Kerr effect.

\begin{figure}[!h]
\centering
\includegraphics[height=20mm]{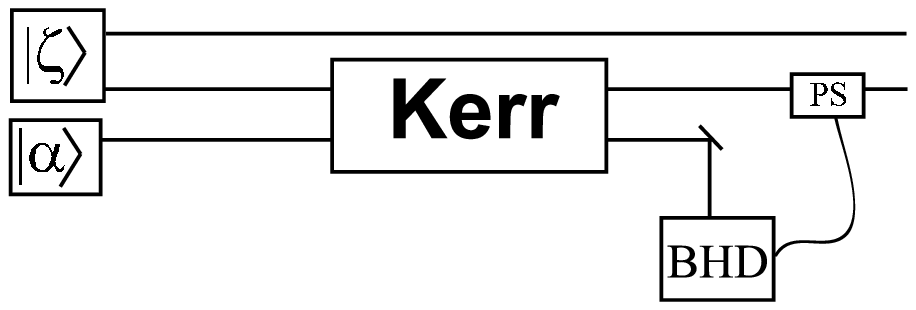}
\caption{\it \small This scheme involves interaction between Bob's beam (one half of $\vert \zeta \rangle$) and an additional coherent state $\vert \alpha \rangle$ via the cross Kerr effect. This is followed by balanced homodyne detection (BHD) and compensating phase shifting (PS). }
\end{figure}

This non-linear interaction is employed for two reasons. Firstly, it is a non-Gaussian operation and is therefore absolutely essential if entanglement concentration is to be achieved. Secondly, this interaction allows the additional coherent state to become entangled with two mode squeezed state (TMSS). This occurs because $\vert \alpha \rangle$ picks up a phase shift $e^{i n \varphi}$ that is dependent on the photon number of Bob's half of the TMSS. In other words, the presence of the TMSS in the non-linear medium causes the coherent state to slow down. The mathematical details of this interaction are encoded in the unitary operator:
\begin{equation}
\hat{U}_{K} = \hat{\mathbb{I}}_{A} \otimes \exp \Bigl( -i \kappa t \hat{b}^{\dagger} \hat{b} \hat{c}^{\dagger} \hat{c} \Bigr),
\end{equation}

\noindent where $\kappa$ is the cross coupling constant of the medium that is related to the $\chi^{(3)}$ nonlinearity. Following \cite{fiurasek:2003}, we define the nonlinear phase $\varphi = - \kappa t$. After the evolution in the Kerr medium, the state is:
\begin{equation}
\vert \psi_{in}(t) \rangle = \sum_{n=0}^{\infty} c_n \vert n,n, \alpha e^{i n \varphi} \rangle.
\end{equation}

It is at this point where we diverge from \cite{fiurasek:2003} and instead of projecting our additional coherent state onto another random coherent state by eight-port homodyne detection, we opt for a single balanced homodyne detection. So, the emerging coherent beam (designated as beam $C$ with mode operators $\{ \hat{c}, \hat{c}^{\dagger} \}$) is then directed into a balanced homodyne detector that measures the quadrature $\hat{x}_{C}^{(\theta)}$. Following the treatment presented in \cite{leonhardt:1997}, the photocurrent is represented by the observable:
\begin{equation}
\hat{I} =  \vert \alpha_{LO} \vert ( e^{i \theta} \hat{c}^{\dagger} + e^{- i \theta} \hat{c} )= \sqrt{2} \vert \alpha_{LO} \vert \hat{x}_{C}^{(\theta)}
\end{equation}

\noindent Note that $\vert \alpha_{LO} \vert$ is the classical amplitude associated to the local oscillator field. Clearly, this observable's eigenstates coincide with the quadrature eigenstates $\vert x_{\theta} \rangle$. Hence, this measurement projects onto one of these states with the probability density:
\begin{equation}
\pi(x_{\theta}) = \mbox{Tr} \bigr ( \hat{\rho}_{C} \vert x_{\theta} \rangle \langle x_{\theta} \vert \bigl).
\end{equation}

\noindent $\hat{\rho}_{C} = \mbox{Tr}_{AB} \bigr (\vert \psi_{in}(t) \rangle \langle  \psi_{in}(t) \vert \bigl)$ is the state of the measurement device with the entangled pair traced out. The whole measurement process is described by:
\begin{eqnarray}
\vert \psi_{in} (t) \rangle \rightarrow \frac{(\hat{\mathbb{I}}_{AB} \otimes \vert x_{\theta} \rangle \langle x_{\theta} \vert )\vert \psi_{in}(t) \rangle}{\sqrt{\pi(x_{\theta})} } = \vert \Psi \rangle \nonumber \\
= (\pi(x_{\theta}))^{-1/2} \sum_{n=0}^{\infty} c_n \langle x_{\theta} \vert \alpha e^{i n \varphi } \rangle \vert n ,n , x_{\theta} \rangle.
\end{eqnarray}

\noindent Here $\langle x_{\theta} \vert \alpha e^{i n \varphi } \rangle $ represents the quadrature wave-function of the coherent state $\vert \alpha e^{i n \varphi} \rangle$, given by \cite{barnett&radmore:1997}:
\begin{eqnarray}
\langle x_{\theta} \vert \alpha e^{i n \varphi} \rangle = \pi^{-1/4} \exp \left ( - \frac{(x_{\theta} - \langle \hat{x}_{C}^{(\theta)} \rangle )^2}{2} \right) \nonumber \\
\times \exp \left( i \langle \hat{x}_{C}^{(\theta+ \pi /2)} \rangle x_{\theta} - \frac{i}{2} \langle \hat{x}_{C}^{(\theta)} \rangle \langle \hat{x}_{C}^{(\theta+ \pi /2)} \rangle \right ),
\end{eqnarray}

\noindent where
\begin{eqnarray}
\langle x_{c}^{(\theta)} \rangle = \sqrt{2} \alpha \cos \left( n \varphi - \theta \right ), \\
\langle x_{c}^{(\theta+ \pi/2)} \rangle = \sqrt{2} \alpha \sin \left( \theta - n \varphi  \right ).
\end{eqnarray}

\noindent At this point, we note that the phase shift attributed to the coherent state by the nonlinear medium is typically very weak. Consequently, $n \varphi$ is very small for all terms $(-\lambda)^n$ that differ significantly from zero as $\lambda <1$ \cite{fiurasek:2003}. This approximation makes it appropriate to use a linear expansion of the sine and cosine terms:
\begin{eqnarray}
\langle x_{c}^{(\theta)} \rangle \approx \sqrt{2} \alpha ( \cos \theta +  n \varphi \sin \theta  ), \\
\langle x_{c}^{(\theta+\pi/2)} \rangle \approx \sqrt{2} \alpha (\sin \theta - n \varphi \cos \theta  ).
\end{eqnarray}

\noindent Subsequently:
\begin{eqnarray}
\langle x_{\theta} \vert \alpha e^{i n \varphi} \rangle \approx \pi^{-1/4}  \exp{ \left( - \frac{(x_{\theta} - \sqrt{2} \alpha \cos \theta )^2}{2} + \beta(x_{\theta}) n \right)} \nonumber \\
\times \exp{ \left ( \frac{i}{2} \left \{ \alpha^2 \sin 2\theta - 2 \sqrt{2} x_{\theta} \alpha \sin \theta + 2 n \gamma(x_{\theta}) \right \} \right )} \nonumber,
\end{eqnarray}

\noindent where
\begin{eqnarray}
\label{beta}
\beta(x_{\theta}) = \sqrt{2} \alpha x_{\theta} \varphi \sin \theta - \alpha^2\varphi \sin2 \theta, \\
\gamma(x_{\theta}) = \sqrt{2} \alpha \varphi x_{\theta} \cos \theta + \alpha^2 \varphi \cos 2 \theta.
\end{eqnarray}

After the measurement, we remove the oscillatory terms $e^{i \gamma(x_{\theta})n }$ and the global phase terms $e^{\frac{i}{2}  (\alpha^2 \sin 2\theta - 2 \sqrt{2} x_{\theta} \alpha \sin \theta)}$ by using a phase shifter as in \cite{fiurasek:2003}. We assume that the properties of the nonlinear medium $\varphi$ and coherent state $\alpha$ are known in advance, possibly through previous experiments and that the result of the quadrature measurement $x_{\theta}$ is stored. This is all the information required to perform this phase shift. Also note that this operation is only possible because the oscillating terms are linear in $n \varphi$ \cite{fiurasek:2003}.

To reveal the form of the output state we calculate the probability density of the quadrature measurement:
\begin{eqnarray}
\pi(x_{\theta}) = \sum_{n=0}^{\infty} c_n^2 \vert \langle x_{\theta} \vert \alpha e^{i n \varphi } \rangle \vert^2 \nonumber \\
=\frac{e^{-(x_{\theta} - \sqrt{2} \alpha \cos \theta )^2} (1- \lambda^2)}{\sqrt{\pi} ( 1 - (e^{ \beta(x_{\theta})}\lambda)^2 )}.
\end{eqnarray}

\noindent Furthermore, if we assume that $\beta(x_{\theta})$ is very small (since $\beta(x_{\theta})$ is proportional to $\varphi$ which is typically very small and $\beta(\sqrt{2} \alpha \cos \theta x_{\theta} )=0$. then a linear approximation is appropriate:
\begin{eqnarray}
\pi(x_{\theta}) =\frac{e^{-(x_{\theta} - \sqrt{2} \alpha \cos \theta )^2} (1- \lambda^2)}{\sqrt{\pi} ( 1 - (1+ \beta(x_{\theta}))^2\lambda^2) }.
\label{assumption}
\end{eqnarray}

\noindent Hence, for small values of $\beta(x_{\theta})$, the probability density is almost Gaussian.

Thus, the output state of the protocol is given by:
\begin{equation}
\vert \psi_{out} \rangle = \sqrt{(1- \lambda'^2)} \sum_{n=0}^{\infty} ( - \lambda' )^{n} \vert n ,n \rangle,
\end{equation}

\noindent where $\lambda' = (1+ \beta(x_{\theta}))\lambda$. The success of the scheme can be examined by comparing the entanglement of the input and output states.

The entanglement content of the input state can be determined by using the {\it Duan} criterion (\ref{duan}) as a measure:
\begin{eqnarray}
\langle (\Delta\hat{U})^2 \rangle +  \langle (\Delta\hat{V})^2 \rangle = \frac{ \Bigl(\vert a \vert - \frac{\lambda}{a} \Bigr) + \Bigl(\frac{1}{a} - \vert a \vert \lambda \Bigr)}{(1 - \lambda^2)}. \nonumber
\label{duan2}
\end{eqnarray}

\noindent For simplicity we choose $a=1$ to give:
\begin{equation}
\langle (\Delta \hat{U})^2 \rangle +  \langle (\Delta \hat{V})^2 \rangle= V_{in}(\lambda) = \frac{2(1- \lambda)^2}{(1- \lambda^2)}.
\label{measure0}
\end{equation}

\noindent The entanglement content of the output state is then:
\begin{equation}
V_{out}((1+\beta(x_{\theta})\lambda) = \frac{2(1 - (1+ \beta(x_{\theta}))\lambda)^2}{(1-((1+ \beta(x_{\theta}))\lambda)^2)}.
\label{measure}
\end{equation}

\noindent Consequently, the entanglement content of the output state is dependent on amplitude of the coherent state $\alpha$, the phase shift generated by the nonlinear medium $\varphi$ and the measurement result $x_{\theta}$. Procrustean entanglement concentration will only occur if $\lambda < (1+\beta(x_{\theta}))\lambda$, which is a condition that can only be satisfied if $\beta(x_{\theta}) > 0$. This success criterion can then be expressed as a condition on the measurement result $x_{\theta}$:
\begin{equation}
x_{\theta}  > \sqrt{2} \alpha \cos \theta.
\end{equation}

\noindent Furthermore, we note that the assumption used to derive (\ref{assumption}) is supported by the {\it Duan} measure since the lower bound of this criterion places an upper bound on the measurement result from the homodyne detection:
\begin{equation}
\langle (\Delta \hat{U})^2 \rangle + \langle (\Delta \hat{V})^2 \rangle \geq 0.
\end{equation}

\noindent Substituting, (\ref{measure}) into the above gives:
\begin{equation}
\beta(x_{\theta}) < 1,
\end{equation}

\noindent which can also be expressed as an upper bound on the possible values of the measurement result $x_{\theta}$:

\begin{equation}
x_{\theta} \leq \frac{(1- \lambda)}{\sqrt{2}\lambda \sin \theta \alpha \varphi}+ \sqrt{2} \alpha \cos \theta = x_{\theta}^{Limit}.
\end{equation}

\noindent Accordingly, the probability of success of this protocol is given by the expression:
\begin{equation}
\mbox{PS} = \int_{\sqrt{2} \alpha \cos \theta}^{x_{\theta}^{Limit}} \frac{e^{-(x_{\theta} - \sqrt{2} \alpha \cos \theta )^2} (1- \lambda^2)}{\sqrt{\pi} ( 1 - ((1+ \beta(x_{\theta}))\lambda)^2)} dx_{\theta}.
\end{equation}

\noindent However, since the probability density is almost a Gaussian and since $\beta(x_{\theta})$ should be greater than zero, then the success probability is approximately $50\%$.

\begin{figure}[!h]
\centering
\includegraphics[height=40mm]{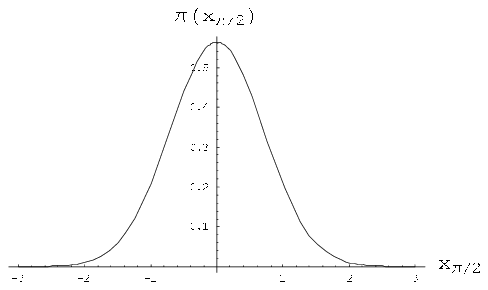}
\caption{\it \small The homodyne measurement probability density is a gaussian centred around the mean value $x_{\theta} = \sqrt{2} \alpha \cos \theta$ in the regime where $\varphi << 1$. In the above $\theta = \pi/2$ and $\alpha = 10^4$ and $\varphi = 10^{-10}$.}
\end{figure}

In the actual implementation of this protocol, the observers Alice and Bob would discard any states which do not obey this condition. Then, they would possess a collection of entangled states with greater entanglement content that the input. However, this collection will not have uniform entanglement since there are many different values of $x_{\theta}$ that satisfy the above condition. Instead, they could only ensure that they had a collection of uniformly entangled states by discarding any states which had a measurement result different from the standard.

\section{Discussion of experimental feasibility}

To discuss the feasibility of our scheme we consider the properties of the probability density (\ref{assumption}) for various values of $x_{\theta}$ corresponding to percentage improvements in the entanglement of the TMSS. Such percentage improvements are calculated from the ratio:
\begin{equation}
\frac{V_{out}((1+\beta(x_{\theta})\lambda)}{V_{in}(\lambda)} = \nu,
\label{ratio}
\end{equation}

\noindent where entanglement concentration occurs only if $\nu <1$ and for a ratio $\nu$ between the variances, we can expect a $(1 - \nu)$$\%$ improvement in entanglement content. In general, the solution of (\ref{ratio}) is given by:
\begin{equation}
\beta(x_{\theta}) = \frac{(\lambda^2-1)(1 -\nu )}{2 \lambda(\lambda(\nu -1) -1 )}.
\end{equation}

\noindent Consequently, we can related the required measurement result $x_{\theta}$ to a desired percentage improvement of entanglement between the two field modes in the TMSS (remembering that $\alpha \in \Re$):
\begin{equation}
x_{\theta} = \frac{(\lambda^2-1)(1 -\nu )}{2 \sqrt{2} \alpha \varphi \sin \theta \lambda(\lambda(\nu -1) -1 )} + \sqrt{2} \alpha \cos \theta.
\label{values}
\end{equation}

\noindent The feasibility of this protocol is then dependent on the probability of obtaining such quadrature values from the measurement. The probability density $\pi(x_{\theta})$ is a gaussian centred at $\sqrt{2} \alpha  \cos \theta$, so the most probable values of $x_{\theta}$ are those where the first term of (\ref{values}) is very small. In other words, for non-trivial probability of success we require:
\begin{equation}
\label{condition}
0 \leq \frac{(\lambda^2-1)(1 -\nu )}{2 \sqrt{2} \alpha \varphi \sin \theta \lambda(\lambda(\nu -1) -1 )} << 1
\end{equation}

\noindent Furthermore, if we regard the non-linear phase shift $\varphi$ and the coherent amplitude $\alpha$ as physical resources in this protocol then it is convenient to express (\ref{condition}) as:
\begin{equation}
\label{condition2}
\alpha \varphi >>\frac{(\lambda^2-1)(1- \nu)}{ 2 \sqrt{2} \sin \theta (\lambda (\nu - 1) -1)}.
\end{equation}

\noindent Using (\ref{condition2}) we can draw a number of conclusions about feasibility. Firstly, the measurement of the phase quadrature $\hat{x}^{(\pi/2)}_C$ is the optimal measurement since $\sin \theta$ is at a maximum when $\theta = \pi /2$.

Secondly, both $\alpha$ and $\varphi$ can be regarded as resources of this protocol and it is possible to compensate for a deficiency in one by strengthening the other. For example, suppose we wish to produce a $10\%$ improvement in the entanglement content of a TMSS with $4.5$ dB squeezing (corresponding to $\lambda =1/2$). Furthermore, if we chose to measure the phase quadrature ($\theta = \pi /2$) then by (\ref{values}), we get:
\begin{equation}
x_{\pi/2} = \frac{0.03}{\alpha \varphi},
\end{equation}

\noindent and the corresponding probability density is:
\begin{equation}
\pi(x_{\pi/2}) \approx 0.6 \exp \left (-\frac{0.0009}{ \alpha^2 \varphi^2}\right ).
\end{equation}

\noindent Now consider how the probability density behaves in different regimes (see figure 3):

\begin{enumerate}
\item When the non-linear coupling between the two radiation modes inside the non-linear medium is weak then $\varphi$ is extremely small. The performance of this protocol is then determined by adjusting $\alpha$ in the ancillary mode. However, the experimentally feasibility of this is questionable. For example, using 1 m of a micro-structured fibre to provide the non-linear interaction and a 10$fs$ pulsed coherent beam with average power 1 mW and repetition rate $80~\mu$Hz for the ancillary state, then it is possible to achieve $\varphi \approx 10^{-9}$ \cite{Agrawal:2001}. To compensate for this tiny non-linearity we would require $\alpha \approx 1.5^7$ which is completely unrealistic with current technology.
\item If we could produce $\varphi \approx 10^{-5}$ then the scheme would be feasible with $\alpha \approx 1.5^3$ .
\item When the nonlinear coupling is strong then the scheme becomes much more viable. For example, if we were able to produce $\varphi \approx 10^{-2}$ then we would only need $\alpha \approx 1.5$ for non-trivial probabilities of achieving a $10\%$ improvement.
\end{enumerate}

\begin{figure}[!h]
\centering
\includegraphics[height=150mm]{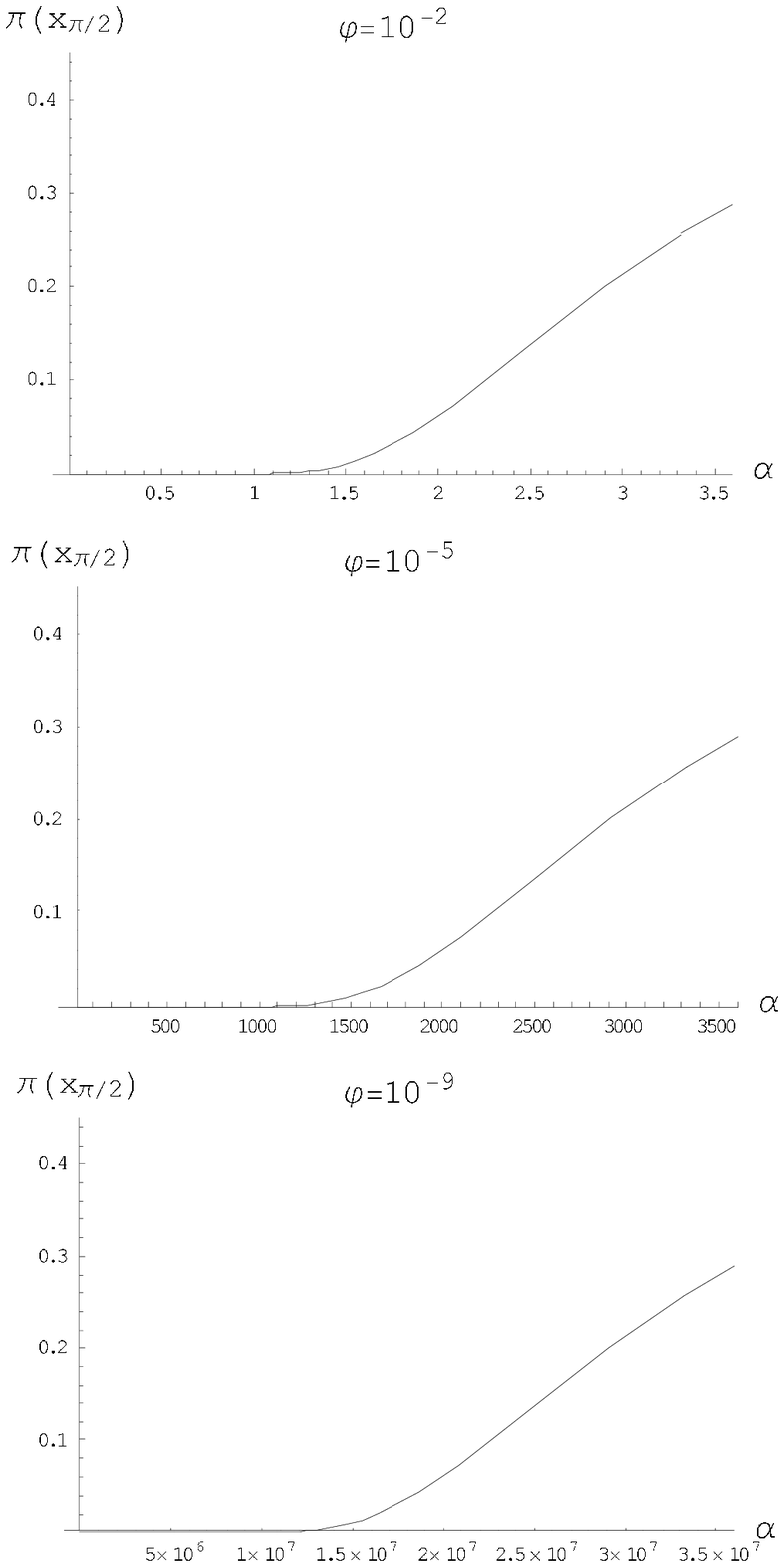}
\caption{\it \small The probability density of achieving a $10\%$ improvement in entanglement of the two mode squeezed vacuum is dependent on both $\alpha$ and $\varphi$. One can compensate for a deficiency in one by strengthening the other, but experimental feasibility is still out of reach with current technology.}
\end{figure}

Thus, with current experimental technology restricted to very small non-linearities, this scheme is not feasible. However, future advancements may offer higher non-linearities in which case our scheme does become viable.

\section{Concluding remarks}

In conclusion, we have shown that Procrustean entanglement concentration can be achieved on a TMSS state by allowing a cross Kerr interaction with an additional coherent state followed by a quadrature measurement on this coherent state. We have found that in replacing the eight-port homodyne measurement suggested in \cite{fiurasek:2003} with balanced homodyne detection, it is possible to have an indication to the success of the entanglement concentration. Indeed, it was demonstrated that entanglement concentration is achieved if $x_{\theta} > \sqrt{2} \alpha \cos \theta$, if this condition is not met then the entanglement content of the TMSS has decreased.

It is useful to compare these two measurement processes to understand why in spite of their differences, they lead to similar results. Firstly, eight-port homodyne or double homodyne detection is designed to extract both total phase and amplitude information from the input state  \cite{paris:2005,leonhardt:1997} and from this information it is possible to construct a coherent state to represent the state of the detector. Thus, double homodyne detection acts to project the input state onto a random eigenstate of the annihilation operator. Due to the non-hermitian nature of the annihilation operator, such a measurement must be modelled by a POVM \cite{paris:2005}:
\begin{equation}
\hat{\Pi}(\alpha) = \frac{1}{\pi} \bigl( \hat{D}(\alpha) \otimes \hat{\mathbb{I}}_{B} \bigr) \vert 1 \rangle \rangle \langle \langle 1 \vert \bigl ( \hat{D}^{\dagger}(\alpha) \otimes \hat{\mathbb{I}}_{B} \bigr ),
\end{equation}

\noindent where $\vert 1 \rangle \rangle = \sum_{n=0}^{\infty} \vert n,n \rangle$ and $\int d^2 \alpha \hat{\Pi}(\alpha) = \hat{\mathbb{I}}$. In contrast, we propose a measurement of an actual observable of the field, namely one of the quadratures of the the auxiliary mode. Such a measurement does not extract all the information from the coherent state, instead only partial information is extracted about both the amplitude and phase and yet, entanglement concentration can still be achieved.

Ultimately, this is because this scheme is dependent on the entangling interaction due to the cross Kerr effect between the squeezed state and the coherent state. The Kerr medium creates entanglement by inducing phase shifts in the coherent state that are dependent on the photon number of the TMSS. Thus, to transfer the entanglement from the coherent state and the TMSS to the two modes of the TMSS is dependent on the the amount of information gained from the measurement about the non-linear phase shifts. Hence, the double homodyne measurement is, in a sense, excessive as it extracts information about both the phase and amplitude of the coherent state when only the former is required. This fact is reinforced in our scheme since, one finds that the optimal measurement for entanglement concentration corresponds to the choice $\theta = \pi/2$ i.e. extraction of the phase information of the coherent state. This insight engenders another, namely that optimal entanglement concentration in our scheme requires a certain harmony between the entangling interaction with the ancillary state and the subsequent measurement on the ancillary.

The general feasibility of our entanglement concentration scheme, and similar schemes, is dictated by the strength of the high order non-linear interaction between the ancillary state and the state of interest. It is clear, that one way to generate these interactions is through the use of non-linear media such as fibres or photonic crystals. However, an alternative method would be the simulation of such effects via measurement induced non-linearities \cite{scheel:2003,filip:2005} where a deterministic entangling interaction could be replace by a conditional effective entangling measurement scheme. It will be the focus of future work to investigate this possibility.

\section{Acknowledgements}

This work was supported by the {\it Engineering and Physical Sciences Research Council} and by the EU project FP6-511004 COVAQIAL. The authors would also like to thank Alvaro Feito-Boirac for useful commments.

\nopagebreak

\bibliographystyle{nature}
\small \small \small
\bibliography{main}

\end{document}